\documentclass[preprint,showpacs,prb,preprintnumbers,amsmath,amssymb,floatfix]{revtex4}
\usepackage{amsmath}
\usepackage{graphicx}% Include figure files
\usepackage{dcolumn}% Align table columns on decimal point
\usepackage{bm}
\usepackage{subfigure}
\usepackage[utf8]{inputenc}
\usepackage{color}
               
\begin{document}

\title{Demixing and confinement in slit pores} %Title of paper

\author{N.G. Almarza, C. Martín, E. Lomba and C. Bores}
\affiliation{Instituto de Química Física Rocasolano, CSIC, Serrano
  119, E-28006 Madrid, Spain}

%\thanks{}

\date{\today}
\begin{abstract}
Using Monte Carlo simulation, we study the influence of geometric
confinement on demixing for a series of
 symmetric non-additive hard spheres mixtures confined in slit pores.
 We consider both a wide range of positive
 non-additivities and a series of pore widths, ranging from the pure
 two dimensional limit to a large pore width where results are
 close to the bulk three dimensional case. Critical
 parameters are extracted by means of finite size analysis. We find
 that for this particular case in which
 demixing is induced by volume effects, phase
 separation is  in most cases somewhat impeded by spatial
 confinement. However, a 
 non-monotonous dependence of the critical pressure and density with pore size is
 found for small non-additivities. In this latter case, it turns out
 that an otherwise stable bulk mixture can be forced to demix by simple
 geometric confinement when the pore width decreases down to
 approximately  one and a half molecular diameters. 
\end{abstract}


\maketitle %\maketitle must follow title, authors, abstract and \pacs

\section{Introduction}
Phase separation under confinement has been for decades a topic of
primary interest both from the technological and fundamental science
standpoints\cite{RPP_1999_62_1573}. It is obvious that the reduction
in the number of neighbors of those molecules adjacent to the pore
walls will induce important phase diagram shifts, whose character will be mostly
dependent on the nature of the wall-fluid (or wall-adsorbate)
interaction. 
In the limit of plain two dimensional confinement the system will exhibit 
bidimensional criticality, which is essentially different -e.g. as critical indices are concerned-
from its bulk three dimensional counterpart\cite{TCritPhen93}. 
We assume that this bidimensional criticality also 
holds for the different levels of confinement  studied in this work.\cite{Binder1992a}

Many new and interesting effects can be induced by confining and the
interplay between adsorbate-adsorbate and adsorbate-pore wall
forces. Very recently, Severin and coworkers\cite{Severin2014} found
evidence of a microphase separation in an otherwise fully miscible
mixture of ethanol and water when adsorbed in a slit pore formed by a
graphene layer deposited on a mica wall. Of utmost interest are also
the effects that confinement have on enhancing or preempting
crystallization of undercooled
fluids\cite{APL_2005_86_103110,JPCM_2006_18_R15}. This has been a key
approach in the attempts to throw some light in the search for the
elusive liquid-liquid critical point in undercooled
water\cite{Biddle2014}, resorting to the preemption of
crystallization induced by tight confinement of water in
nanopores\cite{Chen2006,Bertrand2012} and extensive use of diffraction
experiments in combination with computer simulations. Not long ago,
Fortini and Dijsktra\cite{Fortini2006} explored the possibility of
manipulating colloidal crystal structures by confinement in slit
pores. In contrast, thorough studies on the influence of tunable confinement on
demixing transitions are scarce\cite{Duda2003}. One of the simplest systems that illustrate
demixing in binary mixtures is the non-additive hard sphere system
(NAHS) with positive non-additivity, of which the limiting case of the
Widom-Rowlinson model\cite{Widom1970} has deserved particular
theoretical attention and prompted the development of specially
adapted algorithms to cope with the hard-core singularities and
critical slowing down of the demixing
transition\cite{Johnson1997}. More general instances of the
non-additive hard sphere mixture problem (mostly in the symmetric
case) have been studied in the
two-dimensional limit\cite{Saija2002}, and
in a number of detailed studies in three
dimensions\cite{JCP_1996_104_4180,Gozdz2003,Jagannathan2003,Buhot2005}. 

In this work, we intend to explore thoroughly the demixing transition
of the symmetric non-additive hard sphere mixture under confinement in
a slit pore by means of computer simulation.

The model defined as
mixture of A and B components, is characterized by an interaction
of the type
\begin{equation}
u_{\alpha\beta}(r) = \left\{\begin{array}{ll}
\infty & \mbox{if}\; r \le \sigma (1 + (1-\delta_{\alpha\beta})\Delta)
\\
0 & \mbox{if}\; r > \sigma (1 + (1-\delta_{\alpha\beta})\Delta)
\end{array}\right.
\end{equation}
where $\alpha,\beta$ denote the A and B species, $\delta_{\alpha\beta}$  
is Kronecker's delta,  the
non-additivity parameter is $\Delta > 0$, and $r$ is the interparticle
separation.

We will study a series of confined non-additive hard sphere mixtures
(for various $\Delta>0$ values)  using
extensive semi-grand ensemble Monte Carlo
simulations\cite{Kofke1988,FrenkelSmitbook,Gozdz2003}. The effects of
geometric confinement  are modeled by the presence of hard-core walls,
separated by a distance, $H$, 
that constrain the particle movement in one space direction (along the
$z$-axis as defined here). The fluid particle with thus be subject to
an external potential of the form
\begin{equation}
V^{ext}(z) = \left\{\begin{array}{ll}
0 & {\rm if } \; \sigma/2 \le z \le H - \sigma/2 \\
\infty & {\rm otherwise}.
\end{array}
\right.
\end{equation}
This
%     Cambio x por z (por tradición)   [NGA]
aims at reproducing the behavior of a fluid confined in a slit
pore. Since all interactions at play are purely hard-core, the
demixing transition will result from the interplay of entropic and
enthalpic (i.e. excluded volume) effects. Our calculations range from
the pure two dimensional limit to a relatively large pore width
($10\sigma$, approaching the bulk three dimensional mixture). We have taken advantage of the
particular nature of the interaction to implement cluster
algorithms\cite{PRL_1987_58_86,PRL_1989_62_261,Buhot2005} in order to
cope with the critical slowing down when approaching the consolute
point. Finite size scaling techniques have been applied in order to
provide accurate estimates of the critical points\cite{Gozdz2003}.
% Cambio estimates por points [NGA]
These systems were previously studied by Duda et al.\cite{Duda2003}
by means of mean-field theory and Monte Carlo simulations, considering
two values of the slit width, $H$, and different values of $\Delta$. In most
of the cases they simulate just one system size, corresponding to 
a number of particles $N=1000$. Here we will perform
a comprehensive analysis of the phase diagram for different values of
$H$, and $\Delta$. In addition, for each case several
values of $N$ will be considered, which will allow us to get more reliable estimates of the
phase diagram of these systems, and in particular of the critical points.

The rest of the paper is sketched as follows: in the next section we
briefly summarize the computer simulation techniques we have used, and
our main results are presented together with conclusions and future
prospects in Section III.

\section{Methodology}
\begin{figure}
\includegraphics[width=12cm,clip]{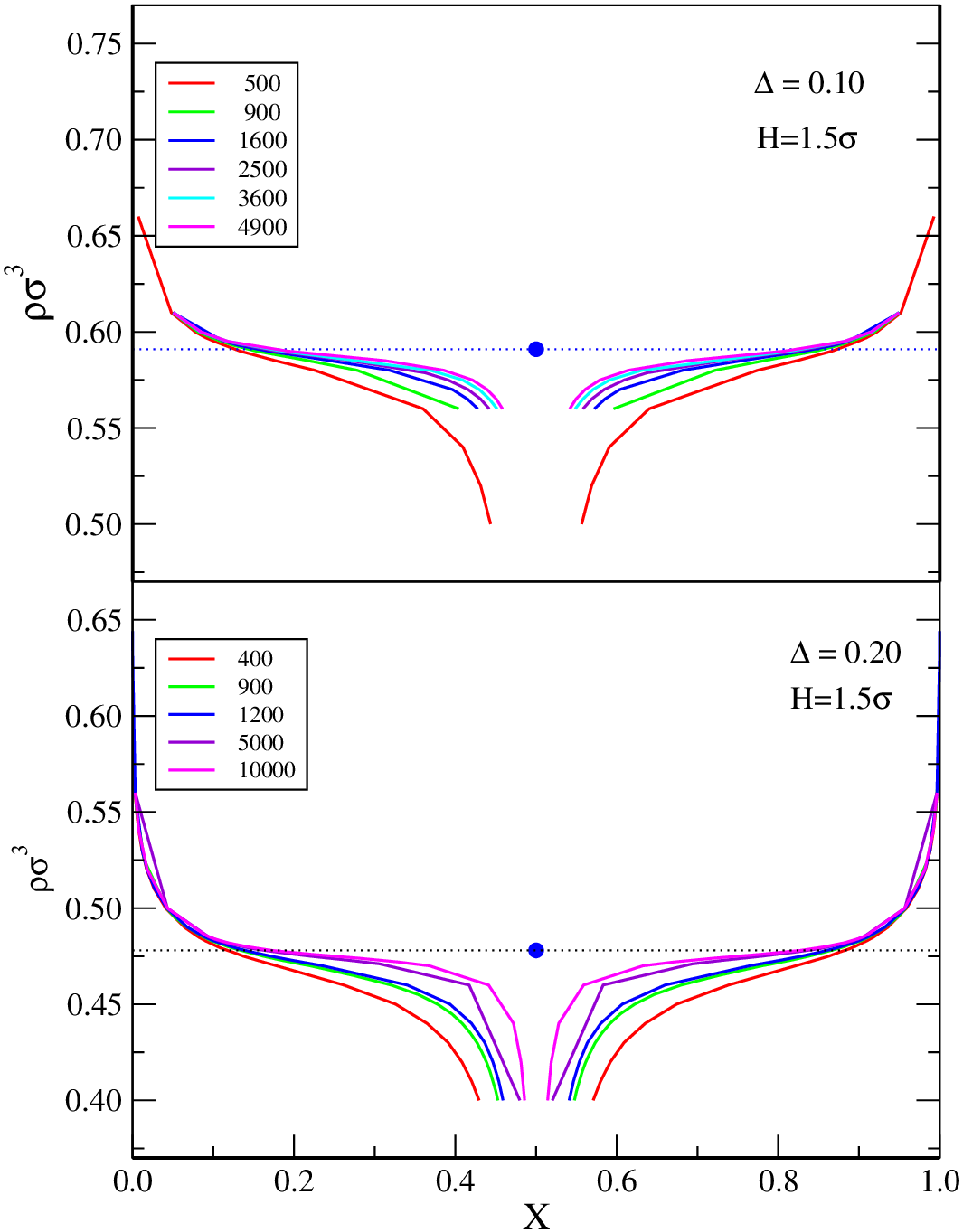}
\caption{Size dependence of phase diagram of symmetric
  non-additive mixtures (with $\Delta=0.1$ and $\Delta=0.2$) confined
  in a slit pore of width $H=1.5\sigma$. The dotted line marks the estimate for the
critical density as obtained from the finite-size scaling analysis.
The symbol correspond to the critical point in the density-mole fraction plane.
}
\label{dphas2}
\end{figure}

Given the particular symmetry of our model, the most appropriate
simulation approach to study the phase equilibria is the use of 
 semi-grand canonical Monte
Carlo (MC) simulations \cite{Kofke1988,FrenkelSmitbook,Gozdz2003}. We impose the
difference between the chemical potentials of the
two components $\Delta \mu \equiv \mu_B - \mu_A$, the volume $V$, the temperature $T$
and keep the total number of particles, $N (= N_{\rm A} + N_{\rm
B})$ fixed; $x = N_{\rm A}/N$ is the concentration of particle
species A. The total number density $\rho = N/V$ is thus fixed. 
In
addition to the conventional MC moves, particles can also modify their identity
(i.e. the species to which they belong)\cite{JCP_1996_104_4180}. The identity
sampling can be performed through an efficient cluster algorithm that involves
all the particles in the systems and that will be presented later in the paper.
After $5
\times 10^5$ MC sweeps for equilibration, our simulations were
typically extended over $2\times 10^6$ MC sweeps to perform
averages.
 A sweep involves $N$ single-particle translation attempts, 
and one cluster move.
 Note that due to symmetry the critical mole fraction of
component $A$ (and $B$) will be $x_c=1/2$, and the demixing transition
will occur at $\Delta \mu=0$.  
When demixing occurs,
 the mole fraction, $X$ 
of the components in the two phases, 
 are computed through
the ensemble averages of the order parameter
\begin{equation} 
\theta=2 x-1,
\label{theta}
\end{equation}
as $X = 1/2 \pm \sqrt{<\theta^2>}/2$. Given the symmetry of the model
and the efficiency of the cluster algorithm, the average of $x$ from the simulations
at $\Delta \mu=0$
will be $<x> \simeq 1/2$, independently of the presence or absence of demixing
at the simulation conditions. By analysis of the mole
fraction histograms for a series of binary mixtures at different
total densities, $\rho = \rho_A+\rho_B$, one can obtain a series of
phase diagrams for each sample size, as illustrated in Figure
\ref{dphas2}, where the extreme size dependence of the results on the
sample size in the neighborhood of the critical point can be readily
appreciated.

It is well known that as the critical point is approached, larger
samples are needed, correlations become long ranged and critical
slowing down must be dealt with somehow. To that aim we have
complemented single particle moves with cluster
moves\cite{PRL_1987_58_86,PRL_1989_62_261,Buhot2005}   following the
Swendseng-Wang strategy\cite{PRL_1987_58_86}. 
Two particles of
the same species are considered linked within the same cluster when their separation is less
than $\sigma (1+\Delta)$.  Note that due to the linking criteria and the hard-core interactions
all the particles belonging to a given cluster are of the same species, and two particles
lying at a distance below $\sigma(1+\Delta)$ are necessarily included in the same cluster.
As a consequence, cluster identity swaps do not lead to particle overlaps, and for the
symmetric case, $\Delta \mu=0$, the procedure leads to a
rejection-free algorithm of composition 
sampling for a fixed set of particle positions. This algorithm rests on two
key elements: (1) Clusters are built following the rules defined
above, and (2) For each cluster one of the two possible identities
($A$ or $B$)  is independently chosen with equal probabilities. 
Along the simulations 
the fraction of 
configurations containing
percolating clusters
 is monitored as an additional signal of the
presence of a phase transition\cite{Stauffer2003}.

Another issue that has to be addressed is the calculation of the
pressure in the confined system with discontinuous interactions. The
scheme proposed by de Miguel and Jackson\cite{Miguel2006} and
further exploited for the Widom-Rowlinson
mixture in Reference \onlinecite{JCP_2007_127_034707} turns out to be the simplest
approach in the present case. In order to estimate the pressure, we perform virtual
compressions of the system (both in the $z$ direction --orthogonal to
the pore walls-- and in the $x,y$ directions). 
The virial pressure is then
computed as
\begin{equation}
\beta P^{z,xy} = \lim_{\Delta V_{z,xy}\to 0} \langle\rho + \frac{1}{\Delta
  V_{z,xy}}n_{o}(\Delta V) \rangle
\label{pres}
\end{equation}
where $\beta= 1/k_BT$ as usual, 
$\Delta V$ is the change of the volume in the compression, $\Delta V = V - V^{\rm test}$,
with $\Delta V >0$,
and $n_{o}(\Delta V)$ is the number of particle
pairs that overlap during the 
virtual (test) compression of the system. 
In practice, the pressure
is calculated by computing $n_{o}$ for a set of values of $\Delta V$
and extrapolating to the the limit  $\Delta V\rightarrow 0$.

Now, the demixing transition is monitored following the evolution and
size dependence of a series of appropriate order parameters. Here we have considered on
one hand, $\theta$, as defined in Eq.~(\ref{theta}), and on the other the
fraction of percolating 
configurations,
 $\chi$. 
A configuration is defined as percolating if (and only if) at least one of its clusters becomes 
of infinite size when considering the periodic boundary conditions; those 
clusters are often denoted as {\it wrapping} clusters. 
With the $\theta$ order
parameter we proceed to perform a Binder cumulant like 
analysis\cite{ZPB_1981_43_119,landau-binder_book_2005}. 
This is done by considering
ratios between momenta of the order parameter probability distribution given as:
\begin{equation}
U_{2n}  = \frac{\langle \theta^{2n} \rangle }{\langle \theta^2 \rangle^{n}},
\label{un}
\end{equation}
where the angular brackets indicate ensemble averages,
and looking at how these quantities vary with the density for different system sizes.
Calculations are carried out for different samples sizes, $N$, and
curves of $\chi$, $U_4$, and $U_6$  are plotted vs. total density
$\rho$. According to the finite size scaling
analysis\cite{landau-binder_book_2005}, the crossing of the curves $U_{2n}(\rho)$ for
different system sizes, 
should define the critical point and be size independent for
sufficiently large samples. In practice, we fit the critical density
estimates, $\rho_c(N)$, obtained from different crossings.
This is done by taking pairs of system sizes, $N_i < N_j$, and
looking for the density $\rho_c(N_j|N_i)$ where the curves of the analyzed 
property for the two system sizes cross.
The results $\rho_c(N_j|N_i)$ for a given $N_i$ are 
taken as estimates for the pseudo-critical
densities for the system size $N_j$, and from then one can extrapolate
the critical density in the thermodynamic limit $(1/N_j \rightarrow 0$).
These extrapolations were done by fitting the results to straight lines of the form\cite{Gozdz2003}
$\rho_c(N) = \rho_c + a N^{-1/(2\nu)}$, where we took $\nu=1$, according to the assumed bidimensional
criticality.
Notice that a more rigorous finite-size scaling analysis  
should be based on 
results  from simulations carried out in either $(N,p,T,\Delta \mu$)
or $(\mu_A,V,T,\Delta\mu)$ ensembles instead of resorting to  $(N,V,T,\Delta\mu)$
semi-grand ensemble simulations.\cite{Fisher1968a,Almarza2012c,Lopez2012b}
The
estimates of $\rho_c$ obtained from the fraction of
percolating configurations and the cumulants are fully consistent
within statistical error bars.
The corresponding
critical pressures are obtained by means of a series of semi-grand canonical
simulations carried out at the critical density and $\Delta \mu=0$ with
varying sample sizes and extrapolating $\beta P^{xy}$ and $\beta P^z$
for $1/N\longrightarrow 0$. An example of the evolution of the order
parameters for the two dimensional limit, (i.e. pore width $H=\sigma$)
and non-additivity $\Delta=0.2$ is presented in Figures \ref{U48} and
\ref{chip}. 

For densities about $\rho_c$, demixing occurs at $\Delta \mu=0$. 
%The mole fraction of the components
%in the two phases $I$, $II$, fulfills: $x_A^{(I)} = x_B^{(II)} = X$, 
%and obviously $x_{A}^{II} = 1 - x_A^{(I)}=1-X$.
The mole fractions of the coexisting phases for each given system size are computed through the order parameter $\theta$,
as: $X_{\pm} = \frac{1}{2} \left[ 1 \pm \sqrt{ \langle \theta^2 \rangle}\right]$.
Using the results for different system sizes, we estimate the composition in the thermodynamic limit
by fitting the results to a second-order polynomial in $(1/N)$.
Then, 
the $X-\rho$ phase diagram can be fully estimated
discarding the equilibrium data close to the critical $\rho_{c}$ (much
affected by sample size dependence), and using the extrapolated data $X(\rho)$,
and a
fit to the approximate\cite{Fisher1968a}  scaling law
\begin{equation}
\left|\frac{X(\rho)-x_c}{x_c}\right| \propto \left|\frac{\rho}{\rho_{c}}-1\right|^\beta,
\label{scal}
\end{equation}
where we assume the system to belong to the two dimensional Ising
universality class\cite{Gozdz2003}, and hence $\beta =1/8$.

\section{Results}
We have considered systems with varying degrees of non-additivity,
ranging from $\Delta = 0.1$ to $\Delta=1$, and pore widths from 
$\sigma$ to $10\sigma$ (see Table
\ref{critres}  for the specific
values). 

\begin{figure}
\includegraphics[width=15cm,clip]{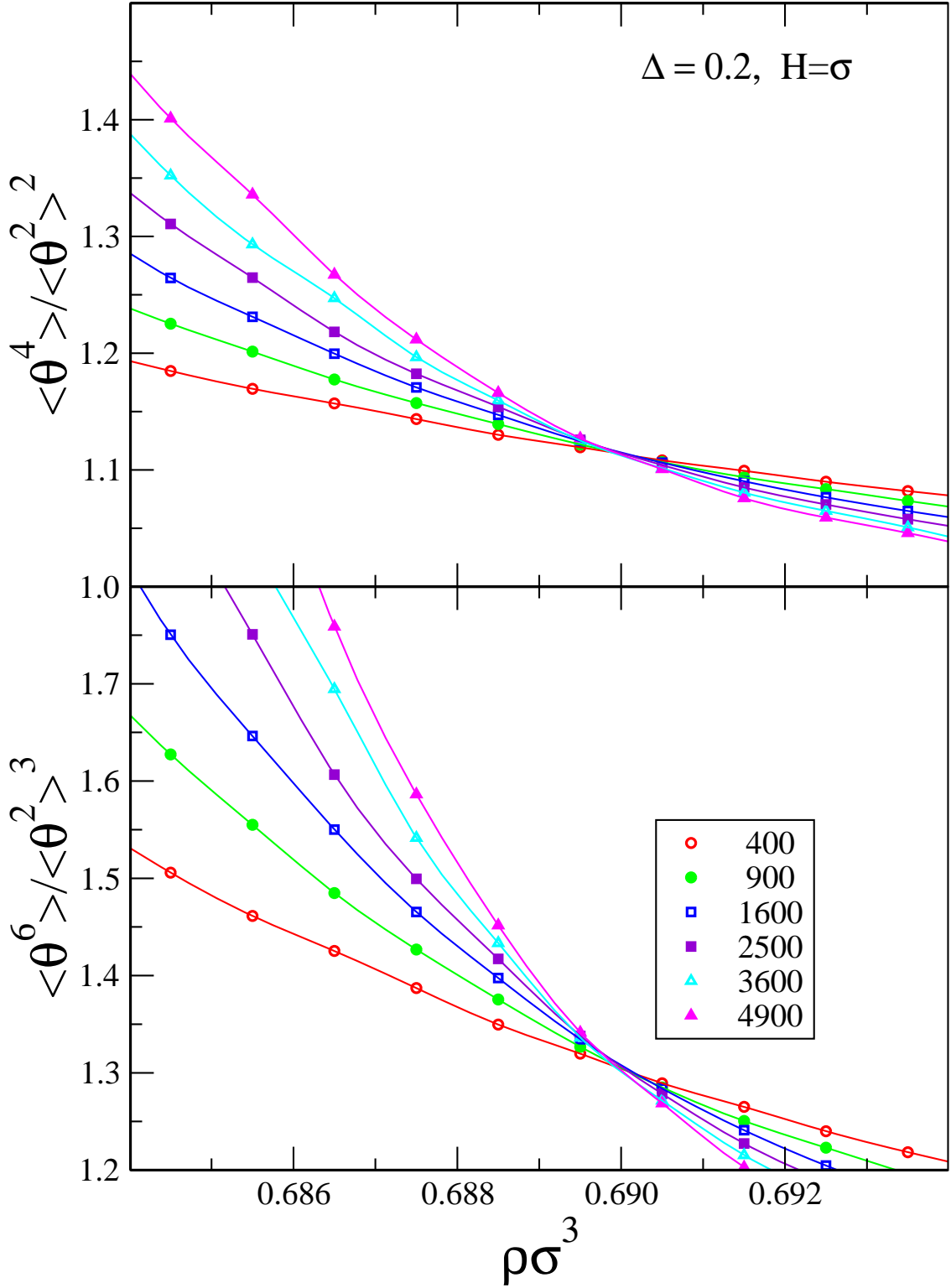}
\caption{Size dependence of the $U_4$, $U_6$ and $U_8$ cumulants of
  the order parameter $\theta$ for the symmetric non-additive hard
  sphere mixture\label{U48}}
\end{figure}

\begin{figure}
\includegraphics[width=15cm,clip]{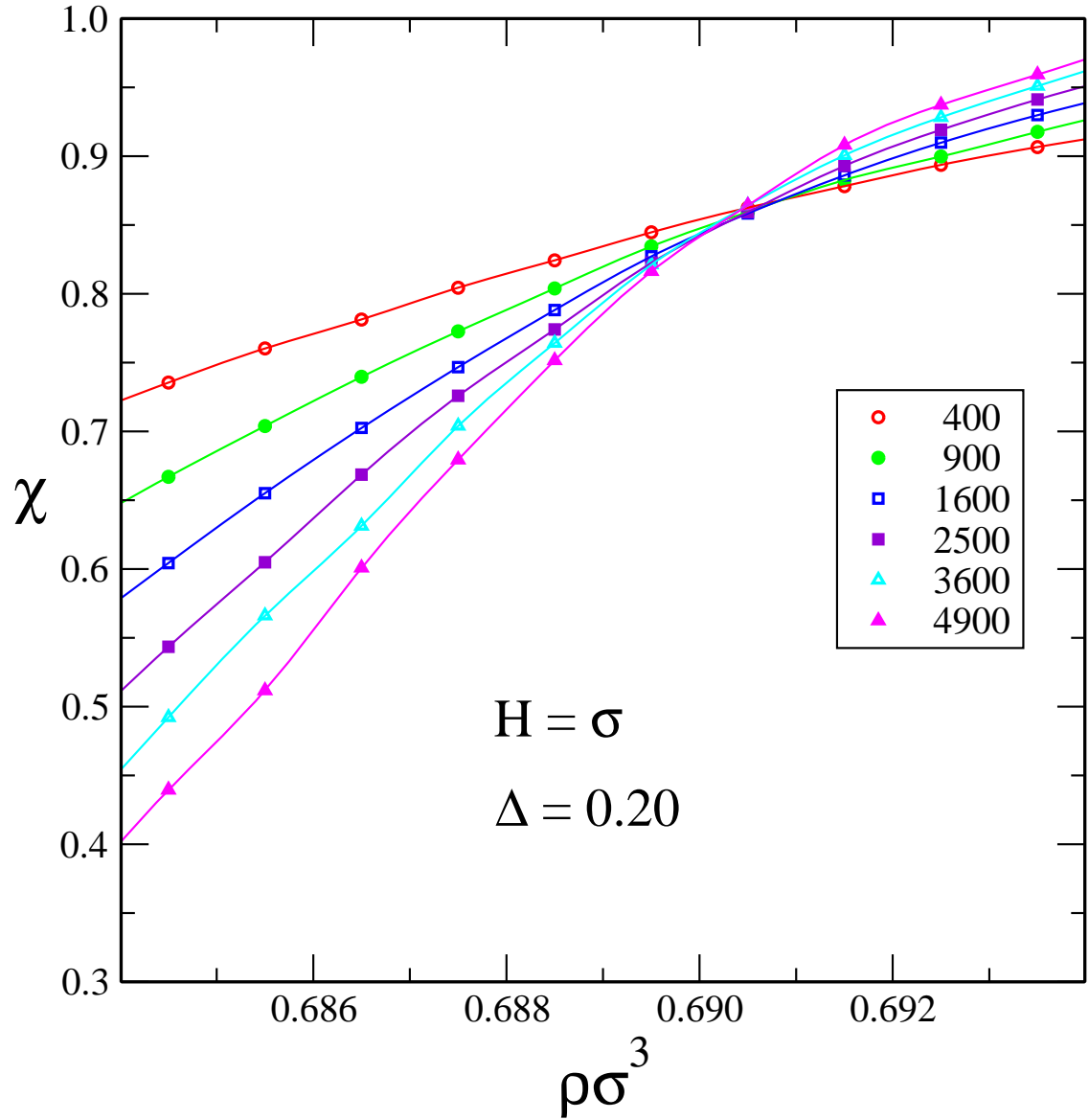}
\caption{Size dependence of the fraction of percolating configurations,
  $\chi$, for the symmetric non-additive hard
  sphere mixture\label{chip}}
\end{figure}

\begin{figure}
\includegraphics[width=15cm,clip]{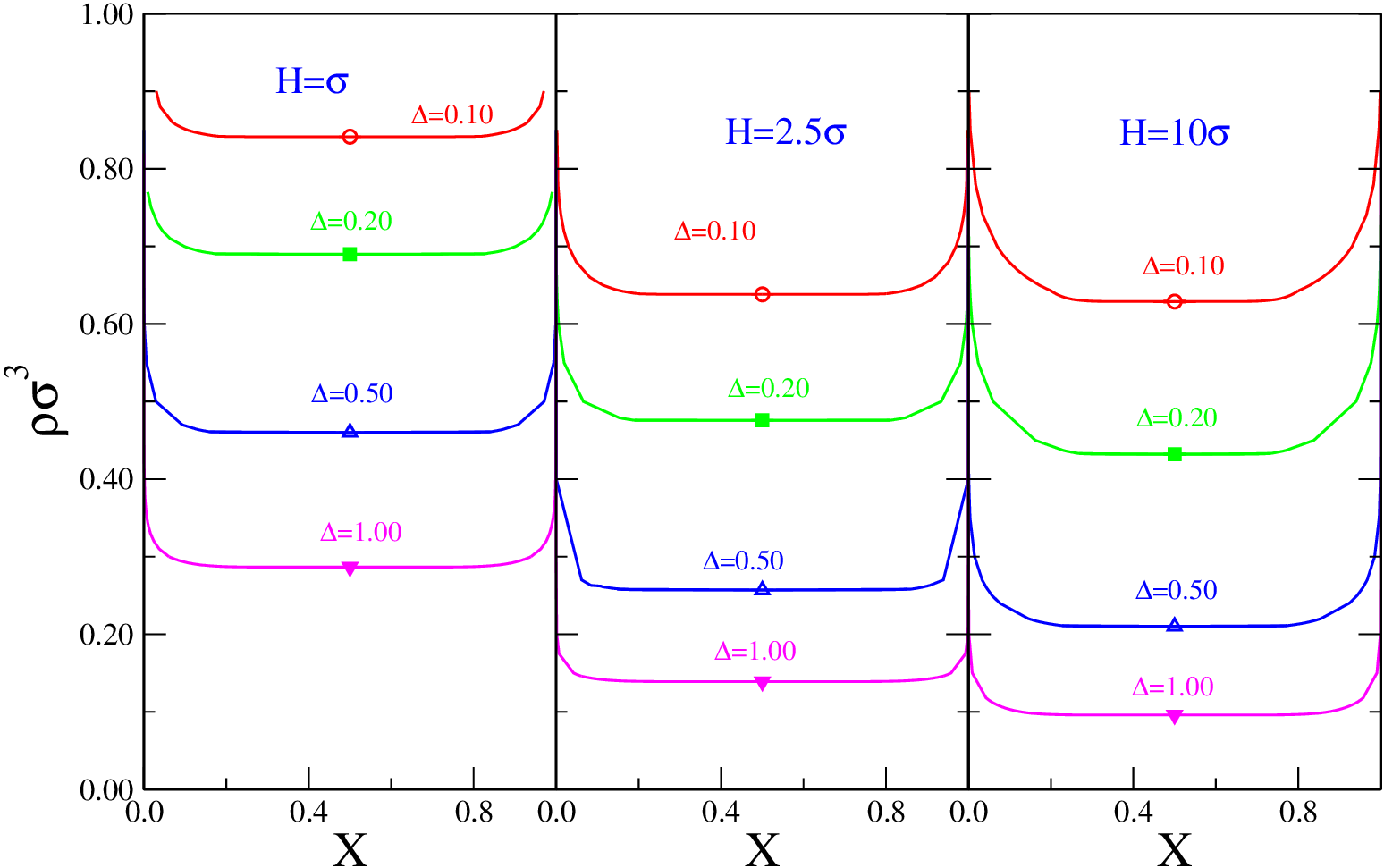}
\caption{Phase diagram of the non-additive hard-sphere mixtures for
  various pore widths and non-additivity parameters \label{DfaseH}}
\end{figure}

Semi-grand ensemble simulations were run for samples of 400, 900,
1600, 2500, 3600 and 4900 particles when $H < 5.5\sigma$. Sample sizes
of 6400 particles were 
included for  pore widths larger than $5.5\sigma$ up to $H=10\sigma$
where an additional sample size of 8100 particles was included. As
mentioned in the previous section, for a given system defined by a
pair ($H,\Delta$), simulations are run for a series of total
densities, $\rho$, and we monitored the behavior of the order parameters
(as illustrated in Figures \ref{U48} and \ref{chip}). Following the
procedures indicated above we obtain a series of phase diagrams as
illustrated in graphs of Figure \ref{DfaseH} for three selected pore
widths, $H=\sigma, 2.5\sigma$, and $10\sigma$. The complete set of
critical properties for most of the systems studied is collected in Table
\ref{critres}.

From Figure \ref{DfaseH} one immediately appreciates
that increasing the non-additivity lowers the critical density,
i.e. favors demixing as expected. In contrast, we observe that
confinement tends to stabilize the mixed phase. This effect is
particularly visible when going from the $H=2.5\sigma$ system to the
two dimensional case, where one sees that the critical density
practically doubles for the two largest non-additivities. Obviously, as the
non-additivity decreases demixing occurs at higher packing fractions
and packing constraints necessarily limit the effects of confinement
on the critical density. Interestingly, we observe that as
$H>2.5\sigma$ the change on the critical density is much smaller, and
practically negligible for the smallest non-additivity. In practice,
as we will see later, for $H=10\sigma$ the critical values of the bulk
three dimensional hard sphere mixture have almost been reproduced. This effect
of stabilization of the mixture due to confinement can be easily
understood when one realizes that the average number of neighbors
is reduced as one goes from the bulk three dimensional system to the two
dimensional one. This implies that particles of a given type A (or B) will have
fewer neighbors of type B (or A) when they are close to the
walls, the limiting case being the two dimensional system. As a
consequence,  these particles will have a lower tendency to demix as
the density (or pressure) is increased. Obviously, the fraction of
particles adjacent to the walls is maximum when $H=\sigma$, and this
fraction decreases rapidly as $H$ increases, and as a consequence the
critical density decreases. Once the pore allows for
two fluid layers inside, the fall in the critical density as the pore
widens is not so pronounced. 

\begin{figure}
\includegraphics[width=15cm,clip]{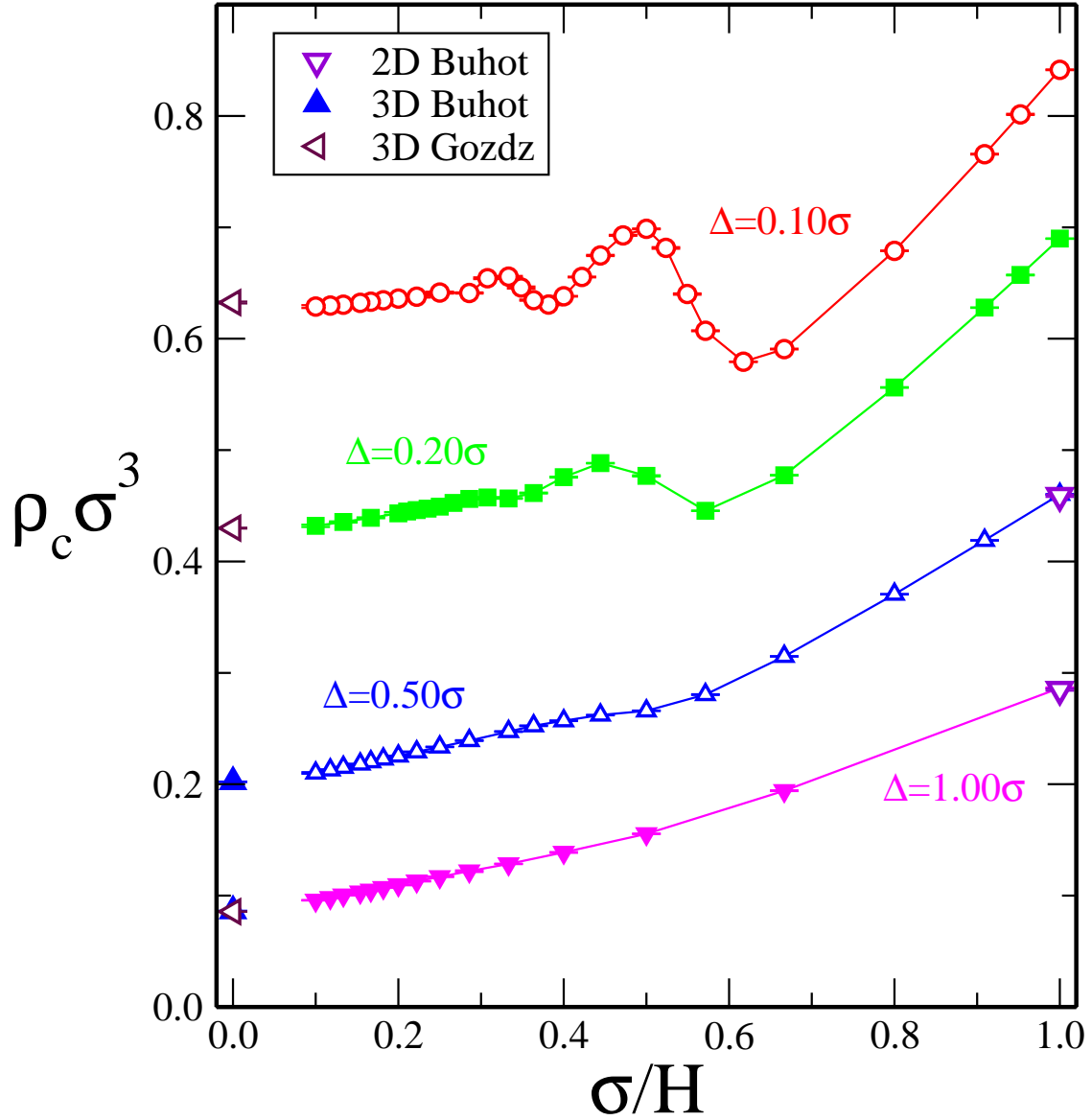}
\caption{Critical density dependence on the slit pore size as
  computed in this work and compared with limiting values in and 2D
  and 3D by  Buhot\cite{Buhot2005} and by Góźdź\cite{Gozdz2003} \label{rhocH}}
\end{figure}

\begin{figure}
\includegraphics[width=15cm,clip]{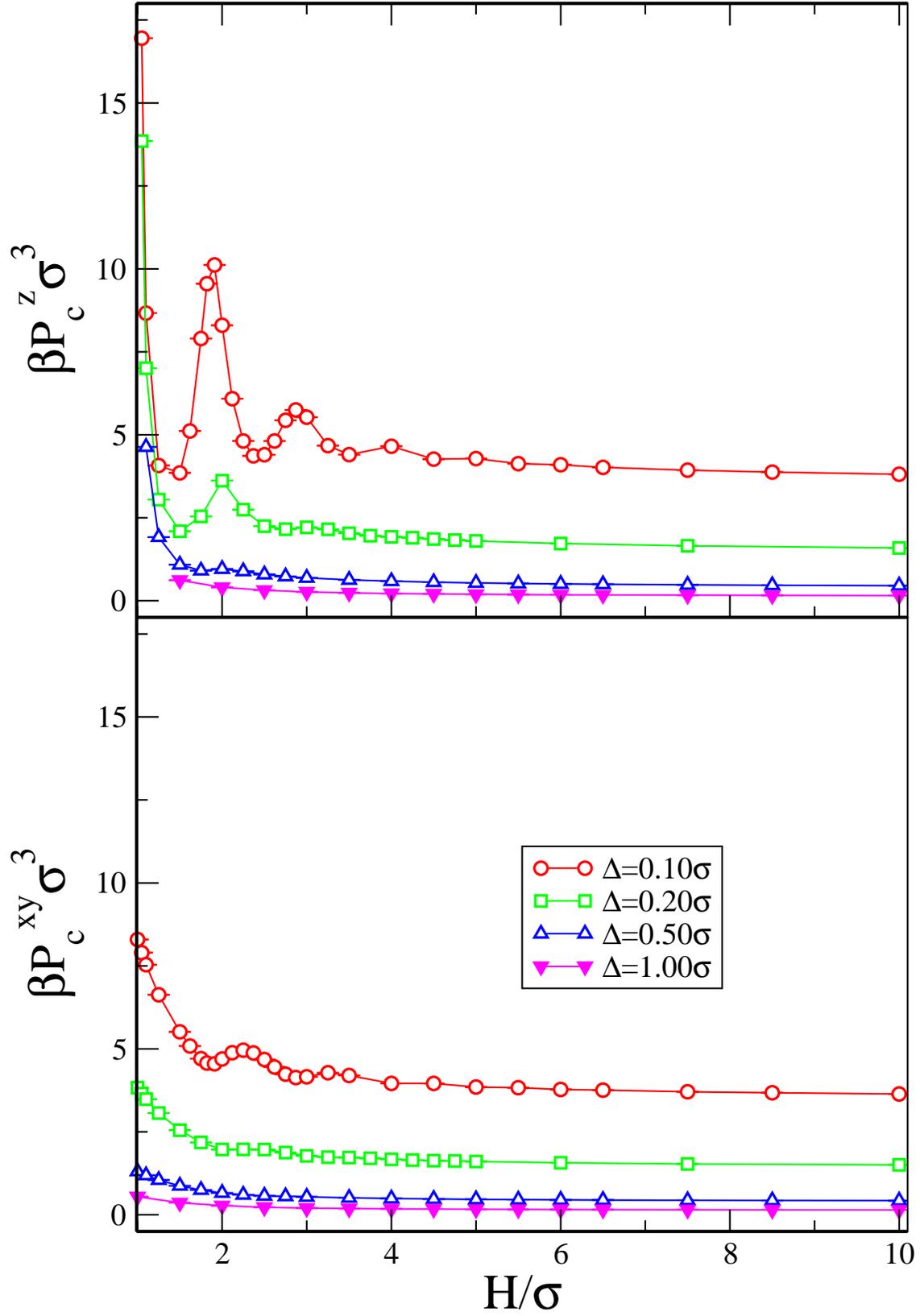}
\caption{Critical pressure dependence on the slit pore size for
  various non-additivity parameters. Upper graph corresponds to the
  pressure on the pore walls and the lower graph to the corresponding
  transverse components\label{PcH}} 
\end{figure}

Now, in Figures \ref{rhocH} and \ref{PcH} we observe the explicit
evolution of the critical density vs $1/H$ and the critical pressure
vs $H$. In Figure \ref{rhocH} some values from the literature for the
two and three dimensional limit are included. As mentioned before, the
critical densities for $H=10\sigma$ practically have already converged
to those of the unconfined system. The dependence of the critical
density on pore size has two linear regimes, which for $\Delta =
1$ and $1/2$ merge continuously at $H \approx
1.5\sigma$. For smaller pore sizes the critical densities grow
rapidly as the pore size shrinks due to the marked decrease in the
number of neighbors induced by the presence of walls. For larger pore
sizes, $H > 4\sigma$, another linear regime with a less pronounced
slope sets in. An interesting feature emerges in the region $1.5\sigma
< H< 4\sigma$ for $\Delta=0.1$ and 0.2: both the critical density and
$\beta P^z_c$ show a clear non-monotonous dependence on H, with maxima
located at $H\approx \sigma, 3\sigma$, and $4\sigma$, --the latter
only visible in the pressure curve-- recalling the neighbor
shell structure of a pair distribution function. We find then that in
the ranges $(n-1/2)\sigma \lesssim H \lesssim n\sigma$ ($n=2,3$) the critical
density and pressure increase (i.e. the mixture is stabilized) when the
pore widens. Note however that on $\beta P^{xy}_c$ the maxima are
shifted towards larger $H$-values, and actually the minima of $\beta
P^{xy}_c$ lie close to the maxima of $\beta P^{z}_c$. Somehow, the
increase in the pressure against the pore walls tends to be compensated
by a decrease of the pressure along the unbound directions.This mismatch
is the obvious result of the lack of isotropy induced by the walls. 

An extreme situation  occurs at $H = 1.5\sigma$ and
$\Delta=0.1$, for which the critical density ($\rho_c=0.591$) is
appreciably lower than that of the bulk\cite{Gozdz2003}
($\rho_c=0.6325(8)$). This actually implies that for certain systems
(i.e. degrees of non-additivity),
a stable mixture can be forced to demix by simple geometric
confinement. In fact one observes
that the maxima in $\rho_c$ -- i.e. local stability maxima for the
mixtures -- occur when the pore can fit approximately an
integer number of layers ($1,2,$ and 3). From these state points,
increasing or decreasing the pore size induces demixing. The effect of
the increase in pore size is easily explained as the result of an
increasing number of neighbors of different species that will prefer
to be in a single component phase. On the other hand, if we focus on the behavior of the
system when going from $H\approx 2\sigma$ to $H\approx 1.5\sigma$, we
realize that  the number of neighbors does not dramatically
change when $H$ varies within these limits, as long as $\Delta$ is
small. In fact, for $\Delta\rightarrow 0$ $H =
(1+\sqrt{2/3})\sigma\approx 1.8\sigma$ the pore still allows for a closed
packed structure of two particle layers with 9 neighbors per
particle, with A and B particles mixed. If the number of neighbors remains approximately constant, the reduction
of available volume  with the shrinkage of the pore width will induce
demixing. The effect will still be present but less patent  when going from $H\approx
3\sigma$ to $H\approx 2\sigma$. Large values of $\Delta$
will destroy this stabilizing effect, e.g. when $\Delta \approx 1/2$ volume
exclusion will prevent the presence of unlike neighbors in adjacent
layers. Small $\Delta$ values allow for this possibility and therefore
higher packing fractions of the stable mixture can be found, by which
the non-monotonous dependence of the critical properties on the pore
width is explained.

In summary, we have presented a detailed study of the effects of
geometric confinement on symmetric mixtures of non-additive hard
spheres. We have found that, as an overall trend, confinement tends to
impede demixing, rising both critical densities and pressures, but
interestingly for small degrees of non-additivity a non-monotonous
dependence is found. In fact, for certain values of the cross
interaction, it is found that confinement can induce demixing by
simple packing effects. In future works, we will address the
effects of competition between energetic and steric contributions to
the intermolecular potential and tunable wall interactions.

\begin{acknowledgments}
The authors  acknowledge the support from the Direcci\'on
General de Investigaci\'on Cient\'{\i}fica  y T\'ecnica under Grants
No. FIS2010-15502 and FIS2013-47350-C5-4-R. The CSIC is also
acknowledged for providing support in the form of the project PIE
201080E120.
% Put your acknowledgments here.
\end{acknowledgments}

%\bibliography{ei}

\newpage

\begin{table}
\caption{Critical parameters for non-additive hard sphere mixtures
  confined in slit pores. Error estimates of 
critical densities and pressures are below the last
significant digits in both instances.}
%
%\tiny
\begin{tabular}{lcccccccccccc}
\hline
 & \multicolumn{3}{c}{$\Delta=0.1$} & \multicolumn{3}{c}{$\Delta=0.2$} & \multicolumn{3}{c}{$\Delta=0.5$} &  \multicolumn{3}{c}{$\Delta=1$}\\
\hline
$H/\sigma$ & $\rho_c\sigma^3$ & $\beta P_{c}^{xy}\sigma^3$ & $\beta P_{c}^z\sigma^3$ & $\rho_c\sigma^3$ & $\beta P_{c}^{xy}\sigma^3$ & $\beta P_{c}^z\sigma^3$ & $\rho_c\sigma^3$ & $\beta P_{c}^{xy}\sigma^3$ & $\beta P_{c}^z\sigma^3$ & $\rho_c\sigma^3$ & $\beta P_{c}^{xy}\sigma^3$ & $\beta P_{c}^z\sigma^3$ \\
\hline
   1.00 &  0.841 &  8.30 &  ---  & 0.690  & 3.84 &  --  & 0.460 & 1.314 & --  &  0.286 &  0.547 &   --         \\
   1.05 &  0.802 &  7.91 & 16.95 &  0.657 & 3.66 & 13.85& --    &  --   &  --& -- &-- &  --    \\
   1.10 &  0.765 &  7.55 &  8.67 &  0.628 & 3.49 & 7.01 & 0.419 & 1.193 & 4.64 & -- & -- & --  \\
   1.25 &  0.679 &  6.64 &  4.08 &  0.555 & 3.07 & 3.05 & 0.370 & 1.051 & 1.919& -- & --& --  \\
   1.50 &  0.591 &  5.52 &  3.86 &  0.477 & 2.56 & 2.10 & 0.314 &  0.876 &  1.089 & 0.194 &0.365 & 0.621 \\
   1.75 &  0.607 &  4.70 &  7.91 &  0.445 & 2.18 & 2.54 & 0.280 & 0.751  & 0.912 &   --    &  --   &  --      \\
   2.00 &  0.699 &  4.71 &  8.31 &  0.477 & 1.97 & 3.62 & 0.266 &  0.660 &  0.959&  0.155 &  0.275 &  0.405 \\
   2.25 &  0.675 &  4.97 &  4.83 &  0.488 & 1.97 & 2.75 & 0.262 & 0.604 & 0.889   &  -- &  -- &  --    \\
   2.50 &  0.638 &  4.69 &  4.41 &  0.476 & 1.97 & 2.25 & 0.257 &  0.575 &  0.795 & 0.138 & 0.227 &  0.319 \\
   2.75 &  0.635 &  4.24&  5.44 &  0.461 &  1.88 & 2.16 & 0.252 & 0.558  &  0.735 &    --          &       --        &       --        \\
3.00 & 0.656 & 4.16& 5.54& 0.456 & 1.78 & 2.21 & 0.247&  0.542& 0.693& 0.128&  0.202 &  0.268 \\
3.50 & 0.641 & 4.20& 4.40& 0.456 & 1.73 & 2.04 & 0.239 & 0.512 & 0.632& 0.122& 0.189 &  0.241 \\
4.00 & 0.642 & 3.97& 4.66& 0.449 & 1.67 & 1.93 & 0.233 & 0.491 & 0.592 & 0.117 & 0.180 & 0.221 \\
4.50 & 0.638 & 3.96& 4.27& 0.446 & 1.63 & 1.86 & 0.229 & 0.478 & 0.562 & 0.113 &  0.173 &  0.207 \\
5.00 & 0.636 & 3.85& 4.29& 0.443 & 1.61 & 1.80 & 0.225 & 0.468 & 0.541 & 0.109 & 0.167 & 0.196 \\
6.00 & 0.634 & 3.78& 4.11& 0.439 & 1.57 & 1.73 & 0.220 & 0.452 & 0.508 & 0.105&  0.160 & 0.182 \\
7.50 & 0.631 & 3.72& 3.94& 0.436 & 1.54 & 1.66 & 0.215 & 0.440 & 0.482 & 0.100 & 0.153 & 0.169 \\
10.00& 0.630 & 3.65& 3.82 & 0.432& 1.51 & 1.59 & 0.210 & 0.428 & 0.458 & 0.096 & 0.147 & 0.158 \\
\hline
\end{tabular}
\label{critres}
\end{table}

\newpage

\end{document}